\documentclass[10pt]{amsart}
\usepackage{amssymb}
\usepackage{enumerate}
\usepackage[dvips]{graphicx}
\DeclareGraphicsExtensions{.eps,.art,.ART,.ps}
\newcommand{\bbR}{\mathbb{R}}      % real numbers %
\newcommand{\bbN}{\mathbb{N}}      % natural numbers %                     
\newcommand{\cC}{\mathcal{C}}      % causality surface %
\newcommand{\cB}{\mathcal{B}}      % causality ball %
\newcommand{\link}{\operatorname{link}}     % linking number %
\newcommand{\wind}{\operatorname{wind}}     % winding number %                           
\newtheorem{Theo}{Theorem}[section]
\newtheorem{Prop}[Theo]{Proposition}

\newtheorem{Cor}[Theo]{Corollary}
\theoremstyle{definition}
\newtheorem{Def}[Theo]{Definition}
\theoremstyle{remark}

\begin{document}
\begin{abstract}
Given a $(d+1)$-dimensional spacetime $(M,g)$, one can consider the set $N$ of all its null geodesics. If $(M,g)$ is globally hyperbolic then this set is naturally a smooth $(2d-1)$-manifold. The \emph{sky} of an event $x \in M$ is the subset
\[
X = \left\{ \gamma \in N:x \in \gamma \right\}
\]
and is an embedded submanifold of $N$ diffeomorphic to $S^{d-1}$. Low conjectured that if $d=2$ then $x,y \in M$ are causally related \emph{iff} $X,Y \subset N$ are linked (in an appropriate sense). We prove Low's conjecture for a (large) class of static spacetimes. 
\end{abstract}
%
%%%%%%%%%%%%%%%%%%%%%%%%%%%% Definition of title page %%%%%%%%%%%%%%%%%%%%%%%%%%%
%
\title{Linking and causality in $(2+1)$-dimensional static spacetimes}
\author{Jos\'{e} Nat\'{a}rio}
\address{Mathematical Institute, Oxford}
\curraddr{Department of Mathematics, Instituto Superior T\'{e}cnico, Portugal}
%\subjclass[2000]{Primary 53078}
\thanks{This work was partially supported by FCT (Portugal) through programs PRAXIS XXI and POCTI}
\maketitle
%
%
%%%%%%%%%%%%%%%%%%%%%%%%%%%%%%%%%%% Section 1 %%%%%%%%%%%%%%%%%%%%%%%%%%%%%%%%%%
%
\section{Introduction}
In this section we give an overview of the general problem, and describe what is known in the $d=2$ case. Most of this material was first discussed by Low (see \cite{L88}, \cite{L89} \cite{L90a}, \cite{L90b}, \cite{L94}), and can be found in greater detail in \cite{N00}.

Let $(M,g)$ be a $(d+1)$-dimensional globally hyperbolic spacetime with a fixed time orientation, and consider the set $N$ of all its null geodesics (we define a null vector to be a \emph{nonzero} vector $v$ such that $g(v,v)=0$, and hence the constant geodesic is \emph{not} a null geodesic). If $\Sigma \subset M$ is a Cauchy surface, then every null geodesic intersects $\Sigma$ exactly once. On the other hand, at any event $x \in \Sigma$ two future-pointing null vectors are initial conditions for the same null geodesic \emph{iff} they are linearly dependent, or, equivalently, \emph{iff} their orthogonal projections on $T_x\Sigma$ are linearly dependent. Consequently, $N$ can be identified with the tangent sphere bundle $TS\Sigma$ (recall that $\Sigma$ endowed with the metric induced by $g$ is a Riemannian manifold of dimension $d$). We use this fact to define a differentiable structure on $N$, and notice that, due to smooth dependence of the solutions of the geodesic equation on its initial data, this structure is independent of the choice of Cauchy surface. Thus the set of all null geodesics of a globally hyperbolic $(d+1)$-dimensional spacetime is a differentiable manifold of dimension $d+(d-1)=2d-1$, which we call its \emph{manifold of light rays}.

The \emph{sky} of an event $x \in M$ is the subset
\[
X = \left\{ \gamma \in N:x \in \gamma \right\} \subset N.
\]

If $x \in \Sigma$ then $X$ is a fibre of the tangent sphere bundle $TS\Sigma$, and therefore is an embedded submanifold of $N$ diffeomorphic to $S^{d-1}$. Since we are free to regard $N$ as a fibre bundle over \emph{any} Cauchy surface, and every event in $M$ belongs to \emph{some} Cauchy surface, we see that any sky is an embedded $S^{d-1}$.

Let us now assume that $M$ is orientable. Then by choosing a global time function $t:M \to \bbR$ and using the globally defined nonvanishing future-pointing timelike vector field $\frac{\partial}{\partial t}$ we can orient all Cauchy surfaces $\{t=\text{constant}\}$; hence we can define an orientation on each sky by orienting tangent spheres on each Cauchy surface (it is easy to check that this orientation does not depend on the choice of global time function).

If $x,y \in M$ are not in the same null geodesic then $X \sqcup Y$ is a \emph{smooth link}, i.e., a disjoint union of embedded $S^{d-1}$s in a smooth manifold of dimension $2d-1$. Recall that a smooth one-parameter family of diffeomorphisms $\Phi_{t}:[0,1] \times N \to N$ is said to be a \emph{smooth (ambient) isotopy} if $\Phi_{0}$ is the identity map. Two links $X_{1} \sqcup Y_{1}$ and $X_{2} \sqcup Y_{2}$ are said to be \emph{equivalent} if there exists a smooth isotopy $\Phi_{t}:[0,1] \times N \to N$  such that $\Phi_{1}(X_{1})=X_{2}$ and  $\Phi_{1}(Y_{1})=Y_{2}$. It can be shown that two links $X_{1} \sqcup Y_{1}$ and $X_{2} \sqcup Y_{2}$ are equivalent \emph{iff} there exists a \emph{smooth motion} carrying one into the other, i.e., \emph{iff} there exist smooth one-parameter families of embeddings $f_{t}:S^{d-1} \to N$ and $g_{t}:S^{d-1} \to N$ such that $f_0(S^{d-1})=X_1$, $g_0(S^{d-1})=Y_1$, $f_1(S^{d-1})=X_2$, $g_1(S^{d-1})=Y_2$ and $f_t(S^{d-1}) \cap g_t(S^{d-1}) = \varnothing$ for all $t \in [0,1]$.

If $(x_1, y_1)$ and $(x_2, y_2)$ are two pairs of non-causally related events, one can easily construct smooth curves $\alpha,\beta:[0,1] \to M$ such that $\alpha(0)=x_1$,  $\beta(0)=y_1$, $\alpha(1)=x_2$, $\beta(1)=y_2$ and $\alpha(t), \beta(t)$ are not in the same null geodesic for all $t \in [0,1]$. This induces a smooth motion of $X_{1} \sqcup Y_{1}$ into  $X_{2} \sqcup Y_{2}$, which are therefore equivalent. We conclude that the skies of any two non-causally related events belong to the same equivalence class, which we define as the \emph{unlink} in $N$ (notice that although there exists a natural choice for the unlink in  $\bbR^{2d-1}$ or $S^{2d-1}$, this is not the case in general for an arbitrary smooth $(2d-1)$-dimensional manifold).
 
Having said that, we now focus on the case when $d=2$ and $\Sigma$ is diffeomorphic to a subset of $\bbR^2$. In this case, $N$ is diffeomorphic to a subset of the tangent sphere bundle of $\bbR^2$, which in turn is diffeomorphic to the interior of a solid torus in $\bbR^3$. It will prove useful to fix a particular embedding $\sigma:TS\bbR^2 \to \bbR^3$ (which we shall call the {\em standard embedding}). Thus, if $(r,\theta)$ are the usual polar coordinates in $\bbR^2$, and $\varphi$ is the coordinate in the fibres of $TS\bbR^2$ corresponding to the angle with the positive $x^1$-direction, we define
\[
\sigma(r,\theta,\phi) = ((2+\tanh r \cos\theta)\cos\varphi,(2+\tanh r \cos\theta)\sin\varphi, \tanh r \sin \theta)
\]

This particular embedding has the advantage of carrying the unlink in $N$ (as defined above) to the usual unlink in $\bbR^3$. Notice that the skies of events on $\Sigma$ are mapped to circles of constant $(r,\theta)$, which for convenience we call {\em meridians}. From now on we shall identify  $N$ with $\sigma \left( N \right)$. Clearly a smooth motion of a smooth link in $N$ is a smooth motion of this link in $\bbR^3$, and consequently if two links are equivalent in $N$ they must be equivalent in $\bbR^3$.

Recall that in $\bbR^3$ there is a well-known link invariant, namely the {\em linking number}. Given a link $X \sqcup Y$, this is simply the integer given by the {\em Gauss integral},

\[
\link(X,Y)=\frac{1}{4 \pi}\oint_{X} \oint_{Y} \frac{({\bf r}-{\bf s}) \cdot (d{\bf r}\times d{\bf s})}{\left\| {\bf r}-{\bf s} \right\|^3}
\]

\noindent (where ${\bf r},{\bf s}:S^1 \to \bbR^3$ are embeddings for $X,Y$), or, equivalently, the integer such that $\left[ X \right]$ is $\link(X,Y)$ times an appropriate generator of $H_1(\bbR^3 \setminus Y)$, depending on the orientation of $Y$ (see \cite{R90}). In particular if  $X \sqcup Y$ is the unlink then $\link(X,Y)=0$. Notice also that, as is easily seen from the Gauss integral formula, $\link(X,Y)=\link(Y,X)$.

Let $x,y \in M$ be two events not in the same null geodesic. Recall we can always assume that $y \in \Sigma$, i.e., that $Y$ is a meridian. In general, however, $x \not\in \Sigma$. Since we can think of $N$ as $TS\Sigma$, we have a natural smooth projection $\pi:N \to \Sigma$. Clearly, $\pi(X)$ is the intersection of the light cone of $x$ with $\Sigma$, and is a piecewise smooth closed curve not containing $y$; we call this the {\em wavefront} generated by $x$ at $\Sigma$. Since $X$ is an oriented curve, $\pi(X)$ can be given the induced orientation.

We can think of $\pi$ as quotienting $N$ by its meridians, and consequently we can identify $\Sigma$ with the quotient space. Thus any smooth surface on $N$ intersecting each meridian exactly once can be identified with $\Sigma$. In this manner we can think of $y$ and $\pi(X)$ as a point and a curve on $\Sigma \subset N$.
One can always choose $\Sigma \subset N$ such that $\Sigma \cap X$ is finite. It is then possible to construct an isotopy of $N$ deforming $X$ into a curve which approaches (with any required accuracy) the wavefront $\pi(X)$ plus a finite number of meridians, one hanging from each intersection of $X$ and $\Sigma$. Indeed, if $\varphi$ is the angular coordinate along the meridians such that $\Sigma=\{\varphi=0\}$, one has but to consider isotopies of the form $\varphi \mapsto \varphi + \delta(t, \varphi)$, where $\delta:[0,1] \times [0,2 \pi] \to [0, 2 \pi]$ is a nonnegative smooth function vanishing for $t=0$ and $\varphi=0, 2 \pi$ and approaching $2 \pi - \varphi$ from below as $t \to 1$. Since the linking number of any two disjoint meridians is zero, we see that $\link(X,Y)=\wind(\pi(X),y)$, where $\wind(\pi(X),y)$ is the winding number of the curve $\pi(X) \subset \Sigma$ around the point $y \in \Sigma$ (recall that $\Sigma$ is diffeomorphic to a subset of $\bbR^2$).

(This result plus the fact that $\link(X,Y)=\link(Y,X)$ allows us to make the following nontrivial observation: if $x,y$ are not in the same null geodesic and $\Sigma_x, \Sigma_y$ are arbitrary Cauchy surfaces through $x,y$ then the winding number of the wavefront generated by $x$ at $\Sigma_y$ around $y$ is equal to the winding number of the wavefront generated by $y$ at $\Sigma_x$ around $x$).

As an example, consider Minkowski's $(2+1)$-dimensional spacetime, i.e., $\bbR^3$ endowed with the line element $ds^2=dt^2-\left(dx^1\right)^2-\left(dx^2\right)^2$, and take hypersurfaces of constant $t$ as Cauchy surfaces. Then all wavefronts are circles, and all linking numbers are therefore either zero or one. Since an event on the Cauchy surface is causally related to the event generating the wavefront {\em iff} it is either inside the wavefront (in which case the winding number is $1$) or on it, we see that two events in Minkowski $(2+1)$-dimensional spacetime are causally related {\em iff} their skies either intersect or are linked with linking number $1$.

\begin{figure}[h]
\begin{center}    
        %\leavevmode
        \includegraphics[scale=.4]{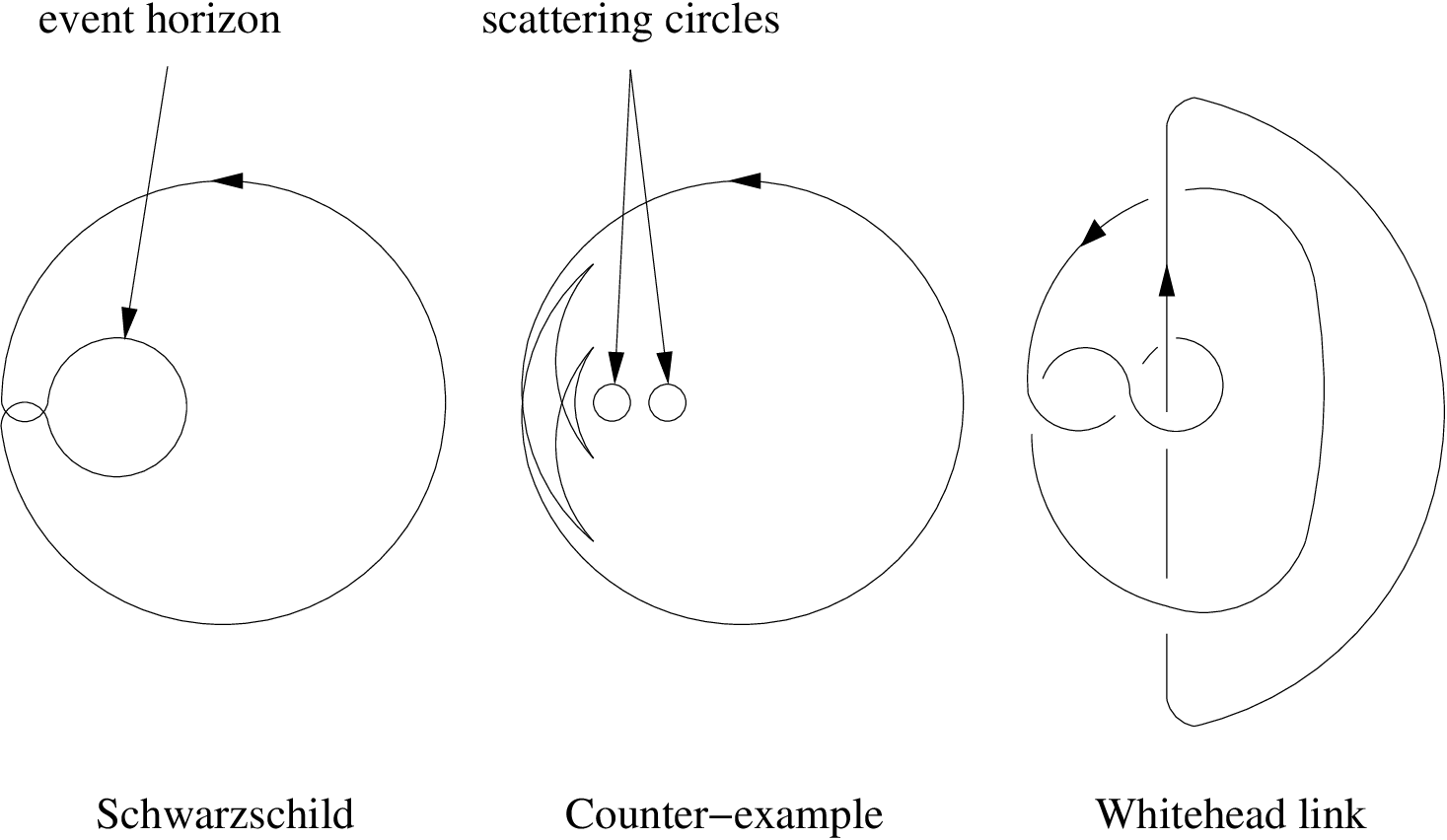}
\end{center}
\caption{}
\label{picture}
\end{figure}

Another example is provided by Schwarzschild's $(2+1)$-dimensional static spacetime, i.e., the region of $\bbR^3$ given in cylindrical coordinates $(t,r,\varphi)$ by $r>1$ endowed with the line element $ds^2=\left(1-\frac{1}{r}\right)dt^2-\left(1-\frac{1}{r}\right)^{-1}dr^2-r^2d\varphi^2$ (we've taken the Schwarzschild radius as our length unit). If we again take hypersurfaces of constant $t$ as Cauchy surfaces, then it is possible to show that the wavefronts are as shown in figure \ref{picture}, wrapping around the event horizon any number of times. It is then easy to see that all winding numbers are either zero or positive, and that an event in the Cauchy surface not on the wavefront is causally related to the event generating the wavefront {\em iff} the winding number is positive. Consequently, two events in Schwarzschild $(2+1)$-dimensional static spacetime are causally related {\em iff} their skies either intersect or are linked with positive linking number (but links do occur with any positive linking number).

It is not true in general that all links formed by the skies of causally related events have nonvanishing linking number. A simple counter-example is provided by $\bbR^3$ endowed with the line element $ds^2=dt^2-\Omega^2\left(x^1,x^2\right)\left(\left(dx^1\right)^2+\left(dx^2\right)^2\right)$, where $\Omega:\bbR^2 \to [1, + \infty)$ is an appropriate smooth function (equal to $1$ except on the circles $\left(x^1\right)^2+\left(x^2\right)^2<1$ or $\left(x^1+4\right)^2+\left(x^2\right)^2<1$, where it increases radially towards the centre). The wavefront of the event $(0,4,0)$ on the Cauchy surface $t=10$ is as depicted in figure \ref{picture}, each pair of cusps corresponding to scattering by one of the circles where the metric is not flat. The appearance of the second pair of cusps allows the existence of events $y$ that, although clearly causally related to $x$, are such that the winding number of $\pi(X)$ around $y$ is zero, and consequently $\link(X,Y)=0$. However $X \sqcup Y$ is {\em not} the unlink. To see that, we notice that the Riemannian metric induced on $\Sigma$ is conformally related to the Euclidean metric, and that we can therefore use the usual angle with the $x$-axis as a coordinate on $TS\Sigma$. To decide on the value of this coordinate along the wavefront, we recall that the tangent vector $\dot{t} \frac{\partial}{\partial t}+\dot{x^1}\frac{\partial}{\partial x^1}+\dot{x^2}\frac{\partial}{\partial x^2}$ to any null geodesic is orthogonal to the light cone of $x$. Consequently, the element of $TS\Sigma$ corresponding to the null geodesic is the normal vector (on $\Sigma$) to the wavefront, oriented so that it points outwards when the wavefront is in the boundary of the causal future. Using these rules one constructs the link shown in figure \ref{picture}, which is the so-called {\em Whitehead link} (and famously {\em not} the unlink).

The above examples have led Low to conjecture that then $x,y \in M$ are causally related \emph{iff} $X,Y \subset N$ are linked. We now proceed to prove Low's conjecture for a (large) class of static spacetimes (which in particular includes the above example). 
%
%
%%%%%%%%%%%%%%%%%%%%%%%%%%%%%%%%%%%%%%% Section 2 %%%%%%%%%%%%%%%%%%%%%%%%%%%%
%
\section{Hyper-regular static $(2+1)$-dimensional spacetimes}
Recall that a $(d+1)$-dimensional spacetime $(M,g)$ is said to be {\em static} if it admits a nonvanishing timelike Killing vector field $T$ whose orthogonal distribution is integrable. The integral submanifolds of the orthogonal distribution endowed with the induced Riemannian metric are isometric Riemannian manifolds (the isometry being generated by the flow of $T$), which we call {\em space manifolds}.  Notice that if $\Sigma$ is any space manifold then the integral lines of $T$ provide a natural projection $p:M \to \Sigma$.
 
\begin{Def}
%\label{regular}
A static $(2+1)$-dimensional spacetime is said to be {\em regular} if
\begin{enumerate}[{\em (i)}]
\item{$(M,g)$ is globally hyperbolic and orientable;}
\item{The space manifolds are Cauchy surfaces diffeomorphic to a subset of $\bbR^2$;}
\end{enumerate}
\end{Def}

From this point on we shall assume that $(M,g)$ is a regular static $(2+1)$-dimensional spacetime. Notice that all three examples considered in the previous section are regular.

Any chart $\left( U, x^i \right)$ \footnote{We shall take latin indices to run from 1 to 2} defined on a space manifold $\Sigma$ can be extended to a chart  $\left( p^{-1}(U), t, x^i \right)$ by using the parameter $t$ along the integral curve of $T$ from the point in $\Sigma$ with coordinates $\left(x^1, x^2 \right)$. In this chart we have $T=\frac{\partial}{\partial t}$. If $h_{ij}$ are the components of the space manifold (Riemannian) metric on the chart $\left( U, x^i \right)$, the line element of the full spacetime manifold on the chart $\left( p^{-1}(U), t, x^i \right)$ is

\[
ds^2=e^{2\phi}dt^2-h_{ij} dx^i dx^j
\]

where

\[
e^{2\phi}=g(T,T)=g\left(\frac{\partial}{\partial t},\frac{\partial}{\partial t}\right).
\]

One has the following {\em Fermat principle} (see \cite{SEF92}):

\begin{Theo}
\label{Fermat}
Let $(M,g)$ be a globally hyperbolic spacetime, $t:M \to \bbR$ a global time function, $x \in M$ an event and $L \subset M$ a smooth timelike curve such that $x \not\in L$. For each future-directed null curve $\gamma \subset M$ through $x$ and $L$ (in that chronological order) let $T(\gamma)=t(y)$, where $y \in L$ is the first intersection of $\gamma$ and $L$ to the future of $x$. Then the null geodesics through $x$ and $L$ are the critical points of $T$.
\end{Theo}

Any segment of null geodesic in the chart $\left( p^{-1}(U), t, x^i \right)$ satisfies

\[
ds^2=0 \Leftrightarrow dt^2=e^{-2\phi}h_{ij} dx^i dx^j.
\]

Using this plus theorem \ref{Fermat} (with the integral curves of $T$ in the role of $L$) it is easy to prove

\begin{Theo}
% \label{lightmetric}
Let $(M,g)$ be a regular static $(2+1)$-dimensional spacetime and $(\Sigma, h)$ a space manifold. Then the projections of null geodesics on $\Sigma$ parametrized by the time coordinate $t$ are the geodesics of the Riemannian metric $e^{-2\phi}h$ on $\Sigma$.
\end{Theo}

For convenience, we make the following

\begin{Def}
Let $(M,g)$ be a regular static $(2+1)$-dimensional spacetime and $(\Sigma, h)$ a space manifold. Then the Riemannian metric $l=e^{-2\phi}h$ on $\Sigma$ is called the {\em light metric}, its geodesics the {\em light geodesics}, the corresponding geodesic spheres the {\em light spheres} and the corresponding geodesic balls the {\em light balls}.
\end{Def}

Clearly light spheres on a space manifold are just the wavefronts generated by various events. Consequently, the light geodesics are perpendicular to the wavefronts (both in the light and space metrics, as they are conformally related). This can also be seen to be a simple consequence of the fact that null geodesics are orthogonal to the null cone, as the tangent vector of the light geodesic is parallel to the orthogonal projection of the tangent vector of the null geodesic on $T\Sigma$.

Since the space manifolds are diffeomorphic to a subset of $\bbR^2$, we can fix global coordinates $\{x^1,x^2\}$ on any particular space manifold $\Sigma$. Again we set the section defined on $TS\Sigma$ by the vector field $\frac{\partial}{\partial x^1}$ as the surface $\{\varphi=0\}$ on $N$, and define the (multivalued) coordinate $\varphi$ in $N$ as the angle with this vector at each point (measured in either metric). Notice that for our purposes it is immaterial whether we use the space metric $h$ or the conformally-related light-metric $l$ in the construction of $TS\Sigma$. For obvious simplicity reasons we shall prefer the latter.

The same nondegenerate light sphere on a space manifold is the wavefront generated by exactly {\em two} events. Indeed, if the centre of the sphere is the point in $\Sigma=\{t=t_0\}$ with coordinates $\left(x^1, x^2\right)$ and its radius is $R>0$, then the sphere is the wavefront generated by the events with coordinates $\left(t_0 \pm R, x^1, x^2\right)$ in the induced chart. Although the skies of these two events project down to the same wavefront on $\Sigma$, they are disjoint curves in $N$, as each point on the wavefront is the projection of two antipodal points in the corresponding fibre of $TS\Sigma$ (one in each sky). Thus any sky is completely defined by its wavefront as long as we indicate a unit normal vector on the wavefront (it suffices to do so at a particular point of the wavefront, as the rest follows from continuity and orthogonality). This extra bit of information is called a {\em coorientation} of the wavefront; it can be thought of as an indication of which way the wavefront is moving. 

Consider the event $x \in M$ with coordinates $\left(t_0-R,x^1,x^2\right)$. For convenience, let us define $x_r \in M$ to be the event with coordinates $(t_0-r,x^1,x^2)$, so that $x_0 \in \Sigma$ and $x_R=x$. It should be clear that an event $y \in \Sigma$ is causally related to $x$ {\em iff} there is a null geodesic through $y$ and $x_r$ for some $0 \leq r \leq R$. Consequently the set of all events in $\Sigma$ which are causally related to $x$ is the union of the light spheres with centre $x_0$ and radius $0 \leq r \leq R$, i.e., is the light ball $\cB$ of radius $R$ and centre $x_0$. $\cB$ can also be thought of as the union of all light geodesic segments with endpoint $x_0$ and length $R$. If one gives the light spheres the obvious coorientation, or, equivalently, the light geodesics the corresponding orientation (away from $x_0$), one can lift $\cB$ onto a surface $\cC$ on $N=TS\Sigma$ by lifting each cooriented light sphere to the corresponding sky, or, equivalently, by lifting each point in the oriented light geodesic to its unit tangent vector.

\begin{Def}
The surface $\cC \subset N$ is called the {\em causality surface} of the event $x \in M$.
\end{Def}

The reason for this name is given in the following

\begin{Prop}
An event $y \in \Sigma$ is causally related to $x$ {\em iff} $Y \cap \cC \neq \varnothing$.
\end{Prop}

\begin{proof}
We've seen that $y \in \Sigma$ is causally related to $x$ {\em iff}
\[
y \in \cB \Leftrightarrow y \in \pi(\cC) \Leftrightarrow \pi^{-1}(y) \cap \cC \neq \varnothing \Leftrightarrow Y \cap \cC \neq \varnothing.
\]
\end{proof}

This is a fundamental step in the proof of Low's conjecture, inasmuch it gives us a precise criterion to decide which skies in $N$ correspond to causally related events. Notice that if one is given a pair of skies it is not obvious in general whether the corresponding events are causally related, even when one of the skies projects down to a point in $\Sigma$, as the wavefront generated by the other event can easily be a very complicated curve.

We now analyse the structure of the causality surface $\cC$ in detail. Since $\cC$ can be seen as a union of skies of events in the line segment $\left \{x_r:r \in [0,R]\right\}$ and one can move from sky to sky along the lifts of light geodesics, it is easy to see that

\begin{Prop}
The causality surface is a immersion of $S^1 \times [0,1]$ in $N$.
\end{Prop}

Consequently the boundary of $\cC$ has two connected components: one is the sky of $x$, and the other is the sky of $x_0$, which is a meridian. For completeness, we identify how $\cC$ can fail to be an embedding:

\begin{Prop}
If the causality surface is not embedded then it self-intersects along segments of lifts of light geodesics. The number of such self-intersection curves is finite if $x_0$ is not conjugate to itself with respect to the light metric.
\end{Prop}

\begin{proof}
If $\cC$ is not an embedded submanifold of $N$, then $X_r \cap X_s \neq \varnothing$ for some $r,s \in [0,R], r < s$ (as $S^1 \times [0,1]$ is compact). Thus $x_r$ and $x_s$ are in the same null geodesic $\gamma$. In other words, $p(\gamma)$ is a light geodesic containing a segment of length $\Delta t=t(x_s)-t(x_r)$ which connects $x_0$ to itself. Consequently, $X_u \cap X_{u+\Delta t}\neq \varnothing$ for $u \in [0,R-\Delta t]$, and $\cC$ self-intersects along a curve. Finally, considering the simultaneous time evolution of the light spheres of centre $x_0$ and radii $0$ and $\Delta t$ it is easy to see that this curve is simply the lift of the segment of light geodesic given by $p\left(\left\{z\in\gamma : \Delta t-R \leq t(z) \leq 0\right\}\right)$. The fact that if $x_0$ is not conjugate to itself there are only finitely many self-intersection curves is a consequence of theorem \ref{conjugate}.
\end{proof}

Recall that $\cC$ is ruled by the lifts of all light geodesics through the point $x_0 \in \Sigma$. Each point in $X_0 \subset TS\Sigma$ is the initial condition for one such geodesic, and since $X_0$ is a meridian it can be parametrized by the coordinate $\varphi$ in $N$. Consequently, an immersion $g: S^1 \times [0,R] \to N$ of $\cC$ is obtained by defining $g(\varphi_0,t)$ to be the unit tangent vector to the light geodesic with initial condition in $X_0$ corresponding to $\varphi_0$ at the point a distance $t$ from $x_0$ along that geodesic (in particular $g\left(S^1 \times \{r\} \right)=X_r$). Also, notice that ${\bf n}=\pi_* g_* \frac{\partial}{\partial t}$ is the unit tangent vector to the light geodesic corresponding to the appropriate value of $\varphi_0$, whereas ${\bf j}=\pi_* g_* \frac{\partial}{\partial \varphi_0}$ is the Jacobi field corresponding to this family of light geodesics. Since ${\bf n} \neq {\bf 0}$ and $h({\bf n},{\bf j})=0$, it is therefore clear that the critical points of $\pi|_\cC$ are the zeroes of ${\bf j}$, i.e., the conjugate points to $x_0$ in the light metric.

We now introduce some useful terminology.

\begin{Def}
A vector $v \in T_{\gamma}N$ is said to be {\em vertical} if $\pi_*v=0$, i.e., if it is a multiple of $\frac{\partial}{\partial \varphi}$. A vertical vector is said to point {\em up} if it is a {\em positive} multiple of $\frac{\partial}{\partial \varphi}$, and to point {\em down} if it is a {\em negative} multiple of $\frac{\partial}{\partial \varphi}$. A vector $v \in T_{\gamma}N$ is said to be {\em horizontal} if it is orthogonal to $\frac{\partial}{\partial \varphi}$ (in the usual Euclidean metric of $\bbR^3 \supset N$).   
\end{Def}

These definitions have obvious extensions to curves and surfaces: a curve or surface will be said to be {\em vertical} at a given point if its tangent space at that point contains a nonzero vertical vector, and will be said to be {\em horizontal} at a given point if all vectors in its tangent space at that point are horizontal. Thus meridians are vertical curves, whereas curves of constant $\varphi$ are horizontal surfaces.

To prove Low's conjecture we just have to show that for every $y \in \Sigma$ such that $Y \cap \cC \neq \varnothing$ and $X \cap Y=\varnothing$ the link $X \sqcup Y$ is not the unlink. We start by showing that we can assume $Y \cap \cC$ to be finite.

\begin{Theo}
\label{conjugate}
If there exist infinitely many light geodesics of length $\leq R \in \bbR^+$ connecting $x_0, y \in \Sigma$ then these points are conjugate.
\end{Theo}

\begin{proof}
Let $\{ \varphi_{0n} \} \in S^1$ be {\em different} initial conditions at $X_0$ for light geodesics connecting $x_0$ to $y$, and $t_n \leq R$ their lengths. Since $S^1 \times [0,R]$ is compact, we can assume that $ \left( \varphi_{0n}, t_n \right) \to \left(\varphi_{0\infty}, t_{\infty} \right)$ as $n \to \infty$ (otherwise extract a convergent subsequence). It is a simple consequence of the continuous dependence of geodesics on their initial conditions that the light geodesic with initial condition $\varphi_{0\infty}$ connects $x_0$ to $y$, and that the light distance between $x_0$ and $y$ measured along this geodesic is $t_{\infty}$. Clearly we can assume $\varphi_{0n}, \varphi_{0\infty}$ to be on the same chart of $S^1$. Let us define
\[
{\bf v}_n=(\varphi_{0n}-\varphi_{0\infty}, t_n-t_{\infty}) \in \bbR^2
\] 
and
\[
{\bf u}_n=\frac{{\bf v}_n}{\left\| {\bf v}_n \right\|}.
\]
(where $\| \cdot \|$ is the usual Euclidean norm in $\bbR^2$). Since ${\bf u}_n \in S^1$, it has at least one sublimit ${\bf u}_{\infty}$. We assume that $\{ \varphi_{0n} \}$ was already chosen so that $\{ {\bf u}_n\}$ is in fact convergent. Setting
\[
\pi(g(\varphi_0,t))=\left(x^1(\varphi_0,t),x^2(\varphi_0,t)\right)
\]
we have from Lagrange's theorem that there exists $\theta_n \in (0,1)$ such that
\[
0=x^1(\varphi_{0n},t_n)-x^1(\varphi_{0\infty},t_{\infty})=\left( \frac{\partial x^1}{\partial \varphi_0}, \frac{\partial x^1}{\partial t} \right) \cdot {\bf v}_n
\]
where the partial derivatives on the right-hand side are calculated at point $(\varphi_{0\infty},t_{\infty})+\theta_n {\bf v}_n$ for some $\theta_n \in (0,1)$. Thus at this point
\[
\left( \frac{\partial x^1}{\partial \varphi_0}, \frac{\partial x^1}{\partial t} \right) \cdot {\bf u}_n=0
\]
and taking the limit we get
\[
\left( \frac{\partial x^1}{\partial \varphi_0}, \frac{\partial x^1}{\partial t} \right) \cdot {\bf u}_{\infty}=0
\]
at point $(\varphi_{0\infty}, t_{\infty})$. The same is true for $x^2$, and we therefore conclude that ${\bf u}_{\infty}$ is in the kernel of $\pi_* g_*$ at $(\varphi_{0\infty}, t_{\infty})$. Consequently,  $x_0$ is conjugate to $\pi(g(\varphi_{0\infty},t_{\infty}))=y$. 
\end{proof}

\begin{Cor}
For every  $\varepsilon > 0$ there exists $z \in B_{\varepsilon}(y) \subset \Sigma$ such that at most finitely many light geodesics connect $x_0$ to $z$.
\end{Cor}

\begin{proof}
We just have to notice that the set of points in $\Sigma$ which are conjugate to $x_0$ has zero Lebesgue measure (see for instance \cite{M73}). 
\end{proof}

\begin{Cor}
There exists $z \in \Sigma$ such that $X \sqcup Z$ is equivalent to $X \sqcup Y$ and $Z \cap \cC$ is finite.
\end{Cor}

\begin{proof}
Since the light sphere $\pi(X)$ is a closed set and $y \not\in \pi(X)$, there exists $\varepsilon >0$ such that $B_\varepsilon (y) \cap \pi(X)=\varnothing$. Any path in $c:[0,1]\to B_\varepsilon(y)\subset\Sigma$ with $c(0)=y$, $c(1)=z$ induces a smooth motion of $X \sqcup Y$ into $X \sqcup Z$. Thus we just have to choose $z \in B_\varepsilon(y)$ such that at most finitely many light geodesics connect $x_0$ to $z$.
\end{proof}

Therefore we shall assume $Y \cap \cC $ to be finite. It is clear from the above that we can assume that $y$ is not conjugate to $x_0$ along any light geodesic, and hence that $Y$ is not tangent to $\cC$ at any of the intersections.

Recall that $\cC$ is ruled by lifts of light geodesics and has $X_0$ and $X$ for boundaries. Consequently, each point of $X$ is connected to a unique point of $X_0$ through the lift of a light geodesic. Since $X_0$ is a meridian, and therefore is a much simpler curve than $X$, a natural idea is to deform $X$ into $X_0$ using the light geodesics connecting them. Each intersection of $Y$ and $\cC$ is then an obstruction to deforming $X \sqcup Y$ into $X_0 \sqcup Y$ (which is the unlink) in this fashion. 

\begin{figure}[h]
\begin{center}    
        %\leavevmode
        \includegraphics[scale=.4]{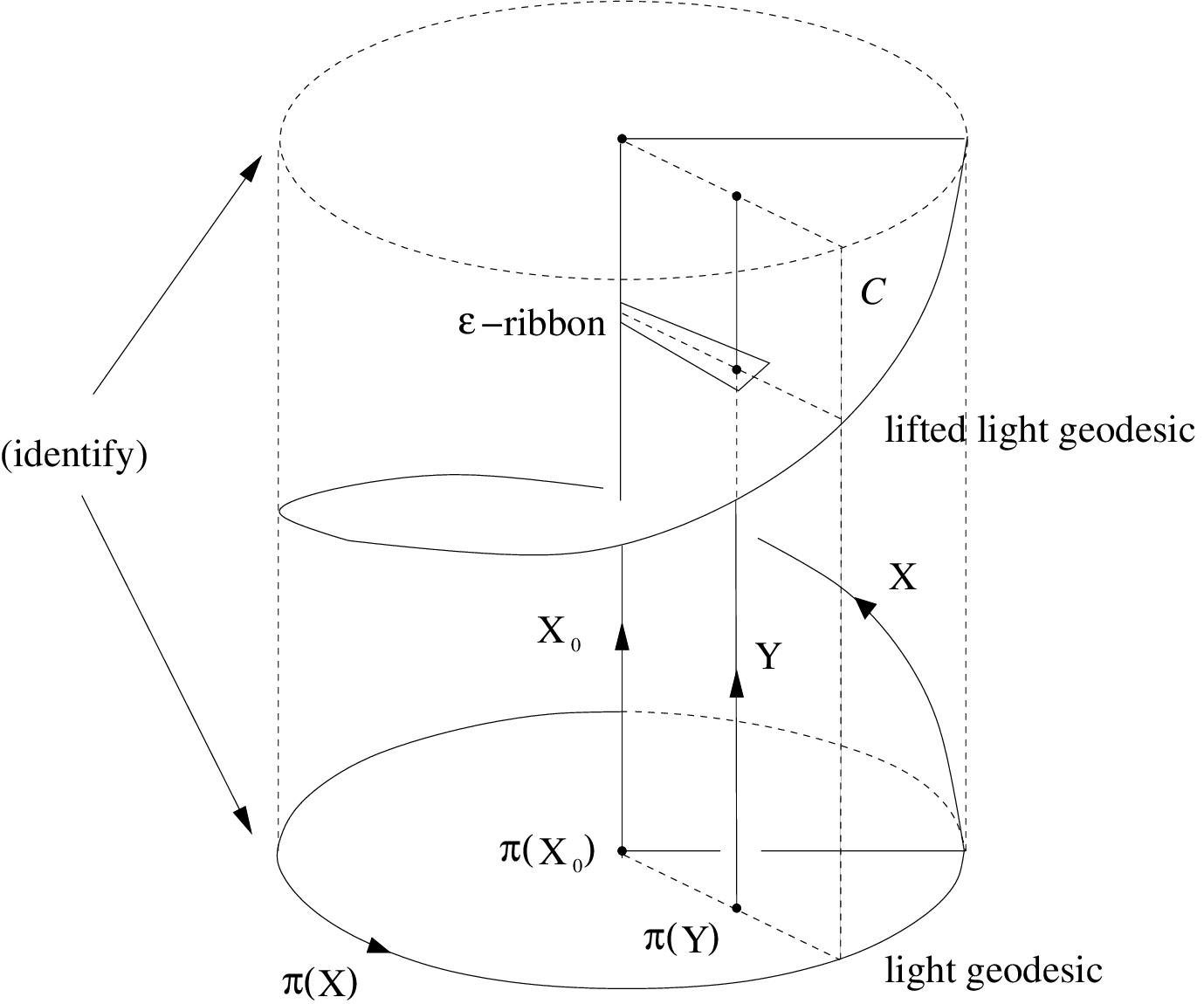}
\end{center}
\caption{}
\label{picture2}
\end{figure}

Let us assume then that $Y \cap \cC \neq \varnothing$, $X\cap Y=\varnothing$. Suppose that $g(\varphi_{0*},t_*) \in Y \cap \cC$ is one of the intersections. Then the $\varepsilon$-{\em ribbon} corresponding to this intersection is the subset
\[
\left\{g(\varphi_0,t):0\leq t\leq t_*+\varepsilon, \varphi_{0*}-\varepsilon \leq \varphi_{0} \leq \varphi_{0*}+\varepsilon \right\} \subset \cC \subset \bbR^3.
\]
(for $\varepsilon>0$). Notice that this set is a (closed) neighbourhood in $\cC$ of the lift of a light geodesic segment joining $x_0$ to $y$. Since the lifts of light geodesics cannot intersect, we conclude that for $\varepsilon$ sufficiently small the (finitely many) $\varepsilon$-ribbons do not intersect. See figure \ref{picture2} for a schematic illustration in the simplest case (Minkowski spacetime).

Let $D\subset\bbR^3$ be a disk with boundary $X_0$ not intersecting $Y$, so that $D \cup \cC$ is a disk with boundary $X$. Let $D_\varepsilon$ be the union of $D$ with all $\varepsilon$-ribbons corresponding to all intersection points in $Y \cap \cC$. Notice that $D \cup \cC$ must always have self-intersections: since the derivative of the geodesic exponential map at the origin is the identity, for small values of $t$ the light geodesics are close to flat metric geodesics, and hence $\pi(X_t)$ must approach a circle with centre $x_0$ and radius $t$; consequently, for $t>0$ small enough, $X_0$ and $X_t$ are linked with linking number $1$, and hence $X_t$ must intersect any disk spanned by $X_0$. However, $D_\varepsilon$ doesn't necessarily have to self-intersect. 

\begin{Def}
A regular static $(2+1)$-dimensional spacetime $(M,g)$ is said to be {\em hyper-regular} if:
\begin{enumerate}[{\em (i)}]
\item
No light geodesic is closed or self-intersects (i.e., all causality surfaces are embedded);
\item
For any two causally related points $x,y \in M$ it is always possible to choose $D$ and $\varepsilon>0$ such that $D_\varepsilon$ does not self-intersect.
\end{enumerate}
\end{Def}

Although this definition may seem somewhat artificial (and indeed is introduced to accommodate the proof of the main result), the class of hyper-regular static $(2+1)$-dimensional spacetimes is fairly large. For instance, all sufficiently small static perturbations of Minkowski spacetime are hyper-regular, as we now show.

\begin{Def}
A curve $c:[0,+\infty) \to \bbR^2$ is said to {\em join} $c(0)$ {\em to infinity} if it is unbounded.
\end{Def}

The set of points in all light geodesics joining two points $\xi, \eta \in \Sigma$ is represented $\Lambda(\xi, \eta)\subset \Sigma$. It is not difficult to see that the following result holds:

\begin{Prop}
If no light geodesic self-intersects and for all $\xi, \eta \in \Sigma \subset \bbR^2$ there exists a curve  $c:[0,+\infty) \to \bbR^2 $ joining $\xi$ to infinity such that  $c\left([0,+\infty)\right) \cap  \Lambda(\xi, \eta) = \left\{ \xi \right\}$ then $(M,g)$ is hyper-regular.
\end{Prop}

This proposition implies in particular that any sufficiently small perturbation of Minkowski spacetime is hyper-regular. Also, it is possible to show that from the three examples in the previous section only Schwarzschild's spacetime is not hyper-regular.

\begin{Prop}
If $(M,g)$ is hyper-regular then for $\varepsilon>0$ small enough, $X \sqcup Y$ can be deformed into $\partial D_\varepsilon \sqcup Y$.
\end{Prop}

\begin{proof}
Since $(M,g)$ is hyper-regular the causality surface is embedded and hence we can deform $X$ into along $\partial D_\varepsilon$ along the lifts of light geodesics without ever intersecting $Y$. 
\end{proof}

We shall assume from this point on that $(M,g)$ is hyper-regular. We see that to understand the link $X \sqcup Y$ we just have to understand the behaviour of the ribbons. Since we can take $\varepsilon >0$ as small as we like, we just have to consider the vector field $g_*\frac{\partial}{\partial \varphi_0}$ along the lift of the corresponding light geodesic, which we do by studying the Jacobi field ${\bf j}$.

At $t=0$ we have ${\bf j}={\bf 0}$, and hence the ribbon starts out vertical. Since the derivative of the geodesic exponential map at the origin is the identity, for small $t$ the light geodesics are close to flat metric geodesics, and hence  ${\bf j}$ points towards the {\em left} with respect to the unit tangent vector ${\bf n}$. This means that the ribbon twists {\em clockwise} around $g_*\frac{\partial}{\partial t}$ as one follows the light geodesic in that direction.

In any time interval $t\in (t_1,t_2)$ such that ${\bf j}\neq {\bf 0}$ we can deform the ribbon so that it stays horizontal (thus initially the ribbon twists by an angle of $\frac{\pi}{2}$ with respect to the vertical direction). However, when a conjugate point is reached this is no longer possible. At such a point one has ${\bf j}={\bf 0}$, but since any Jacobi field satisfies a second order linear differential equation we must have
\[
\nabla_{{\bf n}}{\bf j}\neq {\bf 0}
\]
(where $\nabla$ is the Levi-Civita connection for the light metric). Thus if prior to reaching the conjugate point the Jacobi field points {\em left}, afterwards it must point {\em right}, and vice-versa. Consequently, either ${\bf j}$ or $-{\bf j}$ go from left to right. This means that the corresponding boundary line of the ribbon goes from left to right, and it is easily seen that it does so at smaller values of $\varphi$ than those of the geodesic; in other words, the ribbon twists {\em clockwise} around $g_*\frac{\partial}{\partial t}$ by an angle of $\pi$ at each conjugate point as one follows the light geodesic in that direction (as once it has passed the conjugate point the ribbon can be deformed back into the horizontal position). 

Since we are assuming that $Y$ is not tangent to $\cC$ at the intersections, all ribbons are horizontal at the intersections. Consequently we have proved

\begin{Prop}
As one follows a light geodesic connecting $x_0$ to $y$ in the direction of $g_*\frac{\partial}{\partial t}$, each ribbon rotates clockwise around this vector by an angle of $\left(n+\frac12\right)\pi$ with respect to the vertical direction, where $n$ is the number of conjugate points along the light geodesic.
\end{Prop}

One of the key ingredients in the proof of the main result will be the following trivial observation:

\begin{Cor}
As one follows a light geodesic connecting $y$ to $x_0$ in the direction of $-g_*\frac{\partial}{\partial t}$, each ribbon rotates {\em clockwise} around this vector by an angle of $\left(n+\frac12\right)\pi$ with respect to the vertical direction, where $n$ is the number of conjugate points along the light geodesic.
\end{Cor}

The fact that the rotation is {\em still} clockwise is a consequence of the fact that we have reversed {\em both} the direction in which we are following the curve and the direction of the rotation axis.

Notice that $D_{\varepsilon}$ is a disk, and hence orientable and two-sided. On the other hand, $\partial D_\varepsilon$ has a natural orientation as a deformation of $X$. We can therefore introduce in $D_\varepsilon$ the orientation induced by the orientation of $\partial D_\varepsilon$, and use it to attribute a sign to each of the intersections in $\cC \cap Y = D_\varepsilon \cap Y$. As is well known (see for instance \cite{R90}), $\link(X,Y)=\link(\partial D_\varepsilon, Y)$ is exactly the sum of all signs of all these intersections.

\begin{Theo}
Let $(M,g)$ be a hyper-regular static $(2+1)$-dimensional spacetime, and $x,y \in M$ two causally related events not in the same null geodesic. Then $X \sqcup Y$ is not the unlink.
\end{Theo}

\begin{figure}[h]
\begin{center}    
        %\leavevmode
        \includegraphics[scale=.4]{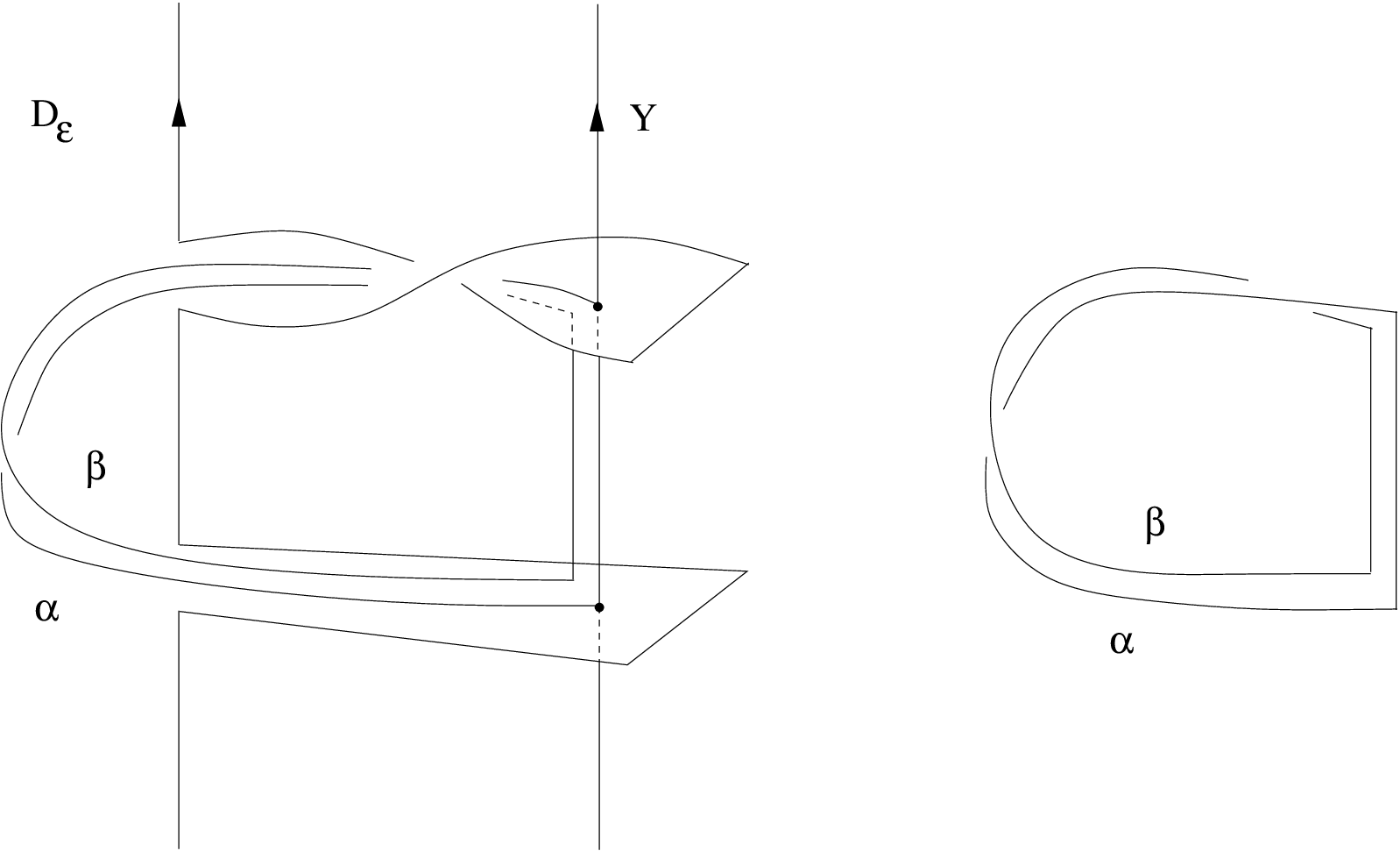}
\end{center}
\caption{}
\label{picture3}
\end{figure}

\begin{proof}
We argue by contradiction: assume that $X \sqcup Y \sim \partial D_\varepsilon \sqcup Y$ is the unlink in $N$; then it is the unlink in $\bbR^3$, i.e., there exists a smooth isotopy carrying $\partial D_\varepsilon \sqcup Y$ to the boundaries of two disjoint disks, say. Consequently there exists a disk $\Delta$ such that $Y=\partial \Delta$ and $\partial D_\varepsilon \cap \Delta=\varnothing$. Since $D_\varepsilon \cap Y \neq \varnothing$, however, we know that  $D_\varepsilon \cap \Delta \neq \varnothing$.

We can assume that  $D_\varepsilon$ and $\Delta$ are transverse (since both $\partial D_\varepsilon$ and $\Delta$ are compact, we just have to deform $\Delta$ in an open neighbourhood of itself not intersecting an open neighbourhood of $\partial D_\varepsilon$ to get rid of any tangency points). Then $D_\varepsilon \cap \Delta$ is a manifold, and hence consists either of closed curves or curves joining two points in the nonempty finite set $D_\varepsilon \cap Y$ (which in particular must have an even number of points).

Take any curve of the latter kind. As a subset of $\Delta$ it joins two points in the boundary, and hence we can turn it into a closed curve $\alpha$ by adding a segment of $Y=\partial\Delta$ between these two points. Since the intersection curves cannot intersect, it is possible to choose the initial curve such that $\alpha$ contains exactly two intersection points in $D_\varepsilon \cap Y$.

Notice that the subset of $\Delta$ bound by $\alpha$ is a disk $\Delta_\alpha$. Consider the curve $\beta$ obtained from $\alpha$ by moving it by a small amount to the interior of $\Delta_\alpha$. The subset of $\Delta_\alpha$ bound by $\beta$ will be another disk $\Delta_\beta$, and $\alpha \cap \Delta_\beta = \varnothing$. Consequently, we must have $\link(\alpha, \beta)=0$.

On the other hand, $\alpha$ is just a curve in $D_\varepsilon$ joining two intersection points at the end of two ribbons plus a segment of $Y$, which is a meridian. The curve $\beta$, in turn, is the push-off of $\alpha$ away from $D_\varepsilon$ along the transverse surface $\Delta$ plus a segment close to the corresponding segment of $Y$ in $\alpha$. This segment cannot wind around $Y$, because $Y \cap \Delta_\beta = \varnothing$ and consequently $\link(Y, \beta)=0$.

Let $n_1, n_2 \in \bbN_0$ be the number of conjugate points along the two light geodesics corresponding to the two ribbons. Since $\alpha \subset \Delta$, we must have $\link(\partial D_\varepsilon , \alpha)=0$; consequently, the intersection points in $\alpha$ must have opposite signs, and hence $n_1+n_2$ is necessarily odd.

As $\beta$ is the push-off of $\alpha$ away from $D_\varepsilon$, $\beta$ must wind around $\alpha$ by an angle of $\left(n_1+n_2+1\right)\pi$ with respect to the vertical direction along the ribbons (see figure \ref{picture3}). Deforming $\alpha$ and $\beta$ along the meridians we then see that we must have $\link(\alpha, \beta)=\frac{n_1+n_2+1}{2}>0$, in contradiction with our previous assertion. We conclude that $X \sqcup Y \sim \partial D_\varepsilon \sqcup Y$ cannot be the unlink.
\end{proof}

%
%
%%%%%%%%%%%%%%%%%%%%%%%%%%%%%%%%%%%%%%% Acknowledgements %%%%%%%%%%%%%%%%%%%%%%%%%%%%
%
\section*{Acknowledgements}

I would like to thank Prof. K. P. Tod for all his help and encouragement; Prof. Sir Roger Penrose for suggesting this problem; and Dr. Marc Lackenby for many useful discussions.

%\bibliographystyle{amsalpha} 
%\bibliography{papel} 
\end{document}